\newif{\ifjournal}
  \journaltrue

\ifjournal
  \documentclass{mn2e}
  \usepackage{graphicx,mathptm}
  \newcommand{\gtrsim}{\ga} 
  \newcommand{\lesssim}{\la} 
\else
  \documentclass{paper}
  \newcommand{\ga}{\gtrsim} 
  \newcommand{\la}{\lesssim} 
\fi

\def\lsim{\mathrel{\rlap{\lower4pt\hbox{\hskip1pt$\sim$}}
    \raise1pt\hbox{$<$}}}                
\def\gsim{\mathrel{\rlap{\lower4pt\hbox{\hskip1pt$\sim$}}
    \raise1pt\hbox{$>$}}}                

\newcommand{\be}{\begin{equation}}
\newcommand{\ee}{\end{equation}}
\newcommand{\bed}{\begin{displaymath}}
\newcommand{\eed}{\end{displaymath}}
\newcommand{\ba}{\begin{eqnarray}}
\newcommand{\ea}{\end{eqnarray}}
 
\newcommand{\chandra}{\emph{Chandra} }

\def\second     {{\prime\prime}}

\begin{document}
\input{psfig.sty} 

\input{psfig.sty} 
\title[Simulating Chandra observations of galaxy clusters]
{Simulating Chandra observations of galaxy clusters}

\author[A. Gardini et al.]
{A. Gardini$^{1,2}$, E. Rasia$^1$\thanks{E-mail: rasia@pd.astro.it},
P. Mazzotta$^{3,4,5}$, G. Tormen$^1$, S. De Grandi$^6$ and L. Moscardini$^7$ 
\\ 
$^1$Dipartimento di Astronomia, Universit\`a di Padova, vicolo
dell'Osservatorio 2, I-35122 Padova, Italy \\ 
$^2$Present address:
Department of Astronomy, University of Illinois, 1002 W. Green Street, 
Urbana, IL 61801, USA\\
$^3$Department of Physics, University of Durham, South Road, 
Durham DH1 3LE, UK \\
$^4$Harvard-Smithsonian Center for Astrophysics, 60 Garden
Street, Cambridge, MA 02138, USA\\ 
$^5$Department of Physics University of Rome "Tor Vergata"
Via Della Ricerca Scientifica, 1 Roma, Italy\\
$^6$INAF-Osservatorio Astronomico di Brera, via Bianchi 46, I-23807 
Merate(LC), Italy\\ 
$^7$Dipartimento di Astronomia, Universit\`a di Bologna, via Ranzani 1, 
I-40127 Bologna, Italy }

\maketitle

\begin{abstract}
Numerical hydro-N-body simulations are very important tools for making
theoretical predictions for the formation and evolution of galaxy
clusters.  Their results show that, accordingly with recent
\chandra and XMM-Newton observations, the atmospheres of clusters of
galaxies have quite complex angular and thermal structures, far from
being spherically symmetric. In many cases the full understanding of
the physical processes behind these features can be only achieved by
direct comparison of observations to hydro-N-body simulations.
Although simple in principle, these comparisons are not always
trivial.  In fact, real data are convolved with the instrument
response and are subject to both instrumental and sky background which
may substantially influence the apparent properties of the studied
features.  To overcome this problem we build the software package
X-MAS (X-ray MAp Simulator) devoted to simulate X-ray observations of
galaxy clusters obtained from hydro-N-body simulations.  One of the
main feature of our program is the ability of generating event files
following the same standards used for real observations.  This implies
that our simulated observations can be analysed in the same way and
with the same tools of real observations.  In this paper we present
how this software package works and discuss its application to the
simulation of \chandra ACIS-S3 observations.  Using the results of
high-resolution hydro-N-body simulations, we generate the \chandra
observations of a number of simulated clusters. We compare some of the
main physical properties of the input data to the ones derived from
simulated observations after performing a standard imaging and
spectral analysis.  We find that, because of the background, the lower
surface brightness spatial substructures, which can be easily
identified in the simulations, are no longer detected in the actual
observations.  Furthermore, we show that, if the thermal structure of
the cluster along a particular line of sight is quite complex, the
projected spectroscopic temperature obtained from the observation is
significantly lower than the emission-weighed value inferred directly
from hydrodynamical simulation.  This implies that much attention must
be paid in the theoretical interpretation of observational
temperatures.
\end{abstract}


\section{Introduction}

Theoretical studies of the dynamical processes underlying the
formation and evolution of galaxies and galaxy clusters are now mainly
based on the results of numerical hydro-N-body simulations.  Since the
pioneering attempts of solving simple N-body problems in the early
'70s, nowadays simulation codes have dramatically improved. Besides
the gas hydrodynamics, their most recent versions may account for some
complex physical processes, that include, but are not limited to, gas
cooling and heating, star formation and feedback, thermal conduction,
magnetic fields (see, e.g., Lewis et al. 2000; Yoshida et al. 2002;
Muanwong et al. 2002; Dolag, Bartelmann \& Lesch 2002; Marri \& White
2003; Kay, Thomas \& Theuns 2003; Springel \& Hernquist 2003;
Tornatore et al. 2003; Borgani et al. 2003).  Simulation results
clearly show that, during their evolution, clusters of galaxies
experience violent events that release an enormous amount of energy in
the intracluster medium (ICM). These events induce strong, but
transient, variation of both ICM density and temperature, so that
their distribution is far to be smooth and spherical symmetric.

From the observational side, thanks to the superb angular and spectral
resolution of the latest generation X-ray satellites, \chandra and
XMM-Newton, we now know that many galaxy clusters, including the ones
previously identified as relaxed, actually present a great deal of
spatial features and have a rather complex thermal structure.  Some of
these include cold fronts (see, e.g., Abell 2142, Markevitch et
al. 2000; Abell 3667, Vikhlinin, Markevitch \& Murray 2001;
RX~J1720+26, Mazzotta et al. 2001; A1795, Markevitch, Vikhlinin \&
Mazzotta 2001; 2A 0335+096, Mazzotta, Edge \& Markevitch 2003), X-ray
cavities (i.e. Hydra A, McNamara et al. 2000; Perseus, Fabian et
al. 2000; Abell 2052, Blanton et al. 2001; Abell 2597, McNamara et
al. 2001; MKW3s, Mazzotta et al. 2002b; RBS797, Schindler et al. 2001;
Abell 2199, Johnstone et al. 2002; Abell 4059, Heinz et al. 2002;
Virgo, Young, Wilson \& Mundell 2002; Centaurus, Sanders \& Fabian
2002; Cygnus A, Smith et al. 2002), X-ray blobs and/or filaments (see,
e.g., Abell 1795, Fabian et al. 2001; Abell 3667, Mazzotta,
Fusco-Femiano \& Vikhlinin 2002a; 2A 0335+096, Mazzotta et al. 2003).
Furthermore, recent observations indicate that also the gas metal
content may have a rather complex distribution that may lead to what
we observe as an off-centre peaked metallicity profile (see, e.g.,
Persueus, Schmidt, Fabian \& Sanders 2002, and Churazov et al. 2003;
Centaurus, Sanders \& Fabian 2002; 2A 0335+096, Mazzotta et al. 2003).
Due to their complex nature and the impossibility of using simple
deprojection techniques, most of these observed features can be
quantitatively studied only through a direct comparison to numerical
simulations.  Ideally, to make these comparisons straightforward one
needs to re-process the simulations themselves through a sort of
virtual observatory in such a way that the information provided is as
much as possible similar to what an observer can obtain through real
X-ray observations of clusters.

In this paper we present X-ray MAp Simulator (X-MAS), a software
package we developed to simulate X-ray observations of galaxy clusters
obtained from hydro-N-body simulations.  The main characteristic of
our code is that, giving as input any hydro-N-body simulation, it
produces as output an event file which is completely similar to what
an X-ray observer would obtain from a real observation.  This means
that the simulated data can be analysed in the same way and using the
same tools of real observations. For the moment our software package
simulates ACIS-S3 \chandra observations only. In the future we will
extend the code to simulate \chandra in the ACIS-I mode and XMM-Newton
observations with both EPIC and MOS detectors.

The outline of the paper is as follow.  In Section~\ref{par:method} we
present the general characteristics of X-MAS.  In
Section~\ref{par:chandra} we show the results of a simulation of an
ACIS-S3 observation of a galaxy cluster.  In Section~\ref{par:dis} we
discuss the possible discrepancy between the projected temperature
derived from the spectral analysis of the observation and the
emission-weighted temperature directly obtained from the hydro-N-body
simulation. In Section~\ref{par:apply} we show some applications of
X-MAS, namely temperature profiles and maps. Finally in
Section~\ref{par:con} we give our conclusions.

\section{X-ray MAp Simulator: The method}\label{par:method}

In this section we describe how the X-MAS package works.  The package
can be divided into two main units.  The first unit is quite general
and does not depend on the specific characteristics of the X-ray
telescope.  For each considered energy channel, it generates a
corresponding map of the differential flux obtained by projecting the
specific emission of each particle along the line of sight.  The
resulting information for the angular position and energy is stored in
a three-dimensional array.  An equivalent way to describe the task of
this first unit is to say that, for each line of sight within a
defined field of view, it calculates and stores the corresponding
projected mass-weighted spectrum.  The second unit takes each spectrum
calculated by the first one and simulates the data relevant to an
observation with a specific X-ray telescope and detector for a defined
amount of time. Of course, this second unit strongly depends on the
characteristics of the X-ray telescope and detector we consider.  At
present our software package simulates \chandra observations in
ACIS-S3 configuration only.  The application of our simulation package
to simulate \chandra in the ACIS-I mode and XMM-Newton observations
with both EPIC and MOS detectors requires an adaptation of this second
unit only.  In the following we describe in details how the two
package units work.

\subsection{First Unit: generating differential flux maps}

The first unit of our package X-MAS requires as input the output of an
hydro-N-body simulation.  After selecting the direction and the depth
of the galaxy cluster for which we want to simulate the observation,
the program generates the cluster projected spectra corresponding to
all the lines of sight in a defined field of view.  This is simply
done by considering an energy interval $[E_{\rm min},E_{\rm max}]$,
which we divide in $n_{_E}$ regular energy channels with energy width
$\Delta E =(E_{\rm max}-E_{\rm min})/n_{_E}$.  The energy interval and
the energy resolution selected for unit one of the program need to be
higher than or equal to the energy response and the energy resolution
of the instrument we intend to simulate later with unit two,
respectively.  We use $E_{\rm min}=0.1$ keV, $E_{\rm max}=10$ keV and
$n_{_E}=495$ (or equivalently $\Delta E =20$ eV\footnote{ The energy
response and resolution of \chandra are [0.1,10.0] keV and $\Delta
E\approx 100$ eV, respectively (see the \chandra Proposers'
Observatory Guide; {\it
http://asc.harvard.edu/udocs/docs/docs.html})}).

For each of these channels we calculate the two-dimensional map of the
corresponding differential flux produced by the simulated galaxy
cluster simply by projecting on a regular grid (1024 pixels $\times$
1024 pixels) the flux $F^{\nu}_i$ of each particle and summing over
all particles.  The differential flux for the cluster is finally
stored in a three-dimensional array in which two dimensions represent
the angular coordinates and the third one is for the energy.

In order to calculate the flux associated with each particle we follow
a procedure very similar to the one described in Mathiesen \& Evrard
(2001).  Starting from its three-dimensional position ${\bf{x}}_{_i}$,
mass $m_i$ and density $\rho_i$, we assign to the $i$-th gas particle
in the simulation an effective volume $V_i=m_i/\rho_i$, which is
assumed for simplicity to be cubic and centred on ${\bf{x}}_{_i}$.

For cosmological sources, the standard relation between flux $F$ and
luminosity $L$ holds: $ F = {L/4\pi d_L^2}$, where $d_L$ is the
luminosity distance (depending on cosmology and on the source redshift
$z$).  Considering differential quantities, the previous relation
becomes $ F_{\nu} = [{(1+z) L_{\nu (1+z)}]/ 4\pi d_L^2}$, where the
${\nu}$ pedex accounts for the dependence on the energy band.  If the
flux is expressed in terms of incoming photons instead of energy, we
can introduce the quantity $ F_{\nu}^{\gamma} \equiv {F_{\nu}/ h\nu}$
(in photon/s/cm$^2$/keV), so that $F_{\nu}^{\gamma} d(h \nu)$
represents the flux of the incoming photons with energy in the range
$[h {\nu}, h ({\nu + d\nu})]$ keV. In general, we prefer to give
quantities like luminosity or flux in terms of incoming photons or
counts (instead of energy) because they can be more easily related to
real observational data. Finally the relation between flux and
luminosity simply becomes
\be 
F_{\nu}^{\gamma} = {(1+z)^2 L_{\nu (1+z)}^{\gamma} \over 4\pi d_L^2}\ .
\ee

The emissivity per unit of frequency $\epsilon_\nu$ in a region of
volume $V$ is related to the luminosity as $ L_\nu = \int_V
\epsilon_\nu \, dV'$; a similar relationship holds for differential
and photon quantities.

For a galaxy cluster the emissivity from a small enough region of
plasma is given by a single temperature thermal model. Given the
electron and hydrogen densities ($n_e$ and $n_{_H}$, respectively),
the emissivity can be written as $ \epsilon = n_e n_{_H}\, P(T,Z)$.
The quantity $P(T,Z)$ is usually called power coefficient and depends
only on the temperature $T$ and metallicity $Z$ of the gas.  Using the
relation above, the photon luminosity can be written as $ L_\nu^\gamma
= \int_V \epsilon_\nu^\gamma \, dV' = P_\nu^\gamma EM$; the quantity $
EM \equiv \int_V n_e n_{_H} dV'$ is often referred to as emission
measure.  The final relation between flux and power coefficients is
then
\be
F_{\nu}^{\gamma} = {(1+z)^2 \over 4\pi d_L^2} EM \, P_{\nu
(1+z)}^{\gamma}(T,Z)\ .
\label{eq:flux}
\ee

Using the temperature and the metallicity of each particle in the
simulation, we calculate $P(T_i,Z_i)$ using the single temperature
thermal model \textsc{mekal} (see, e.g., Kaastra \& Mewe 1993;
Liedahl, Osterheld \& Goldstein 1995, and references therein)
implemented in the utility XSPEC (Arnaud 1996).

To make our simulator of X-ray observation more complete, we allow to
include the effects on the spectra induced by the Galactic HI
absorption.  This is done, at the end of the procedure, by multiplying
the flux in each energy channel by an absorption coefficient given by
the \textsc{wabs} model (Morrison \& McCammon 1983), once a value for
the column density $N_{_H}$ is assumed.

\subsection{Second Unit: Simulating Chandra ACIS-S3 observations}
\label{par:second_unit}

The second unit of our simulation package X-MAS takes as input the
projected spectra produced by the first unit and, after convolving
them with the technical characteristics of a specific instrument,
generates an event file similar to the one obtained from a real
observation.  To do that we use the data simulation command
\textsc{fakeit} in the utility XSPEC (see, e.g., Xspec User's Guide
version 11.2.x; Dorman \& Arnaud
2001\footnote{http://legacy.gsfc.nasa.gov/docs/xanadu/xspec/manual}).
The above command creates simulated data from the input spectral model
by convolving it with the ancillary response files (ARF) and the
redistribution matrix files (RMF), which fully define the response of
the considered instrument, and by adding noise appropriate to the
specified integration time.  Once the data of all spectra have been
simulated, we generate a photon event file satisfying the standards
defined for real observations.  This is quite important because it
allows our mock observations to be analysed by using the same tools
and procedures used for the real ones.

In the following we describe how we use the second unit of our
software package to simulate X-ray observations performed using
\chandra with the back illuminated CCD ACIS-S3.

The ACIS-S3 detector is a square with a grid of 1024 pixels $\times$
1024 pixels, and a field of view of about 8.3 arcmin by side. The
nominal angular resolution is about 0.5 arcsec/pixel.  Although the
instrumental response is position-dependent, it is quite constant over
CCD subregions of $32 \times 32$ pixels, which for convenience we call
detector tiles (or tiles for short).  To fully account for the
instrument response the \chandra calibration team produced 1024 ARF
and 1024 RMF, one for each of the detector tile above.

Detection events have to preserve the spatial and spectral
information.  As explained above, their number and energy are obtained
by executing the command \textsc{fakeit} of the utility XSPEC using
the appropriate ARF and RMF.  To account for the instrumental
background we provide to the \textsc{fakeit} command an appropriate
background file extracted from the blank-sky background dataset
(Markevitch 2001\footnote{http://asc.harvard.edu/ ``Instruments and
Calibration'', ``ACIS'', ``ACIS Background''}) in the same tile
subregion corresponding to the spectrum to be simulated.

To speed up the simulation process we use an adaptive algorithm:
before simulating the data of each spectrum, we estimate the expected
number counts associated with each detector tile.  If this number is
lower than a given threshold we generate the events, otherwise we
iteratively subdivide the region in four squares, until the threshold
is reached.  The spatial position of the events obtained in this way
is then reconstructed by randomly distributing the simulated photons
inside the region by using a weight proportional to the original flux.

\section{ACIS-S3 observation of a simulated cluster}\label{par:chandra}

As a first example of possible applications of our software package,
we generate a 300 ks \chandra ACIS-S3 observation of a high-resolution
hydro-N-body simulation of a galaxy cluster.  The object was selected
from a sample of 17 objects obtained using the technique of
re-simulating at higher resolution a patch of a pre-existing
cosmological simulation. The assumed cosmological framework is a cold
dark matter model in a flat universe with a present matter density
parameter $\Omega_m=0.3$ and a contribution to the density due to the
cosmological constant $\Omega_\Lambda =0.7$; the baryon content
corresponds to $\Omega_B=0.03$; the value of the Hubble constant (in
units of 100 km/s/Mpc) is $h=0.7$, and the power spectrum
normalization is given by $\sigma_8=0.9$.  The re-simulation method,
called ZIC (for Zoomed Initial Conditions), is described in detail in
Tormen, Bouchet \& White (1997), while an extended discussion of the
properties of the whole sample of these simulated clusters is
presented elsewhere (Tormen, Moscardini \& Yoshida 2003; Rasia, Tormen
\& Moscardini 2003).  Here we remind only some of the characteristics
of the cluster used in this paper.  It has been obtained by using the
publicly available code GADGET (Springel, Yoshida \& White 2001);
during the run, starting at redshift $z_{\rm in} = 35$, we took 51
snapshots equally spaced in $\log(1+z)$, from $z=10$ to $z=0$.  Its
virial mass at $z=0$ is $1.46 \times 10^{15} h^{-1} M_\odot$,
corresponding to a virial radius of $2.3h^{-1}$ Mpc; the mass
resolution is $4.5\times 10^9 h^{-1} M_\odot$ per dark particles and
$5\times 10^8 h^{-1} M_\odot$ per gas particles; the total number of
particles found inside the virial radius is 566,374, 48 per cent of
which are gas particles.  The gravitational softening is given by a
$5h^{-1}$ kpc cubic spline smoothing.  Since this particular
simulation does not provide information on the cluster metallicity, we
fixed its value to $Z=0.3\,Z_\odot$.  Furthermore, we assumed a
galactic equivalent column density of $N_H=5\times 10^{20}$ cm$^{-2}$.
Among the available snapshots at different redshifts we chose to
observe the one at $z=0.21$. At this redshift the cluster, having a
virial mass of $1.05 \times 10^{15} h^{-1} M_\odot$ and a virial
radius of $2.2h^{-1}$ Mpc, is undergoing several merger events, so its
structure is quite complex.

\begin{figure}
\psfig{figure=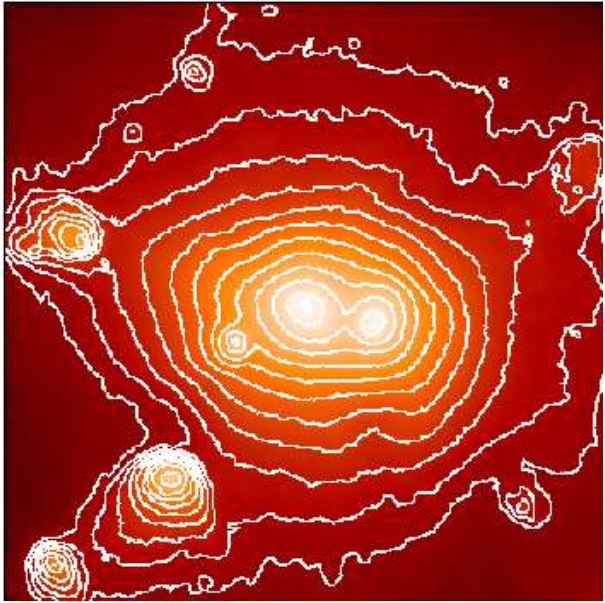,height=8.cm,width=8cm,angle=0}
\caption{Flux map of the simulated galaxy cluster in the [0.1,10.0] keV 
energy range binned to $1\arcsec$ pixels.  The angular size of the map
is 8.3 arcmin, corresponding to approximately 1.7 (proper) Mpc at the
cluster redshift $z= 0.21$.  The superimposed isocontours are obtained
from the image after a Gaussian smoothing with $\sigma= 8^\second$.
Levels are spaced by a factor of $2$ with the highest level
corresponding to $1.8 \times 10^{-15}$ erg cm$^{-2}$ s$^{-1}$.  }
\label{fig:map}
\end{figure}

The cluster flux map, obtained with the first unit of our X-MAS, is
presented in Fig.~\ref{fig:map}. The figure shows the flux in the
[0.1,10.0] keV energy interval.  The displayed region is 8.3 arcmin,
which corresponds to approximately 1.7 (proper) Mpc at the cluster
redshift.  The superimposed isocontours are obtained from the image
after a Gaussian smoothing with $\sigma= 8^\second$.  Levels are
spaced by a factor of $2$ with the highest level corresponding to $1.8
\times 10^{-15}$ erg cm$^{-2}$ s$^{-1}$.  In the external part of the
flux map we find the presence of a number of merging subclumps,
confirming the highly perturbed dynamical phase of the cluster.
Notice that the flux map also shows an orange-skin-like texture
induced by angular structures on scales of the order of few arcsec.
This small-scale structure is actually an artifact that we
deliberately introduced by considering the particle nature of the
Smoothed Particle Hydrodynamics simulation used.  For the purpose of
this paper this artifact turns useful to test the imaging performance
of our simulator on very small scales.

\begin{figure}
\centering  
\psfig{figure=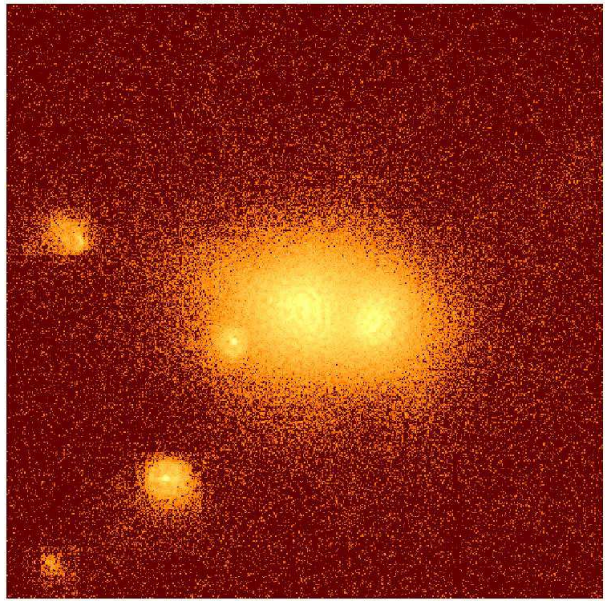,height=8.cm,width=8cm,angle=0}
\caption{Photon image in the [0.3,9.0] keV energy  band 
of the 300 ks ACIS-S3 observation of the same galaxy cluster shown in
Fig.~\ref{fig:map}. The image is background-subtracted,
vignetting-corrected, and binned to $1\arcsec$ pixels.}
\label{fig:events}
\end{figure}

In Fig.~\ref{fig:events} we show the photon image of the 300 ks
\chandra ACIS-S3 observation of the simulated cluster shown in
Fig.~\ref{fig:map}, as obtained by applying the second unit of our
software X-MAS. The image, extracted from the event file in the
[0.3,9.0] keV energy band, is background-subtracted,
vignetting-corrected, and binned to $1\arcsec$ pixels.  We notice
that, as expected, after being observed with {\emph{Chandra}}, the
spatial features present in the simulation, but fainter than the
instrument background, are no longer detected. For example, the three
faint subclumps on the North, North-West and South-West, clearly
visible in Fig.~\ref{fig:map}, have been washed out in
Fig.~\ref{fig:events}.  Conversely all the brighter features appear to
be well reproduced by our simulator.  We find that this is true even
on scales as small as few arcsecs: the previously mentioned angular
artifacts of the simulation on these scales are well visible in the
photon image, indeed.

\section{Spectral analysis: preliminary tests}\label{par:dis}

The spectral analysis of \chandra observations of the simulated
clusters is done by applying standard procedures and tools used for
real observations.  In particular, in order to extract spectra, we use
the \textsc{dmextract} tool of the CIAO software (see
http://cxc.harvard.edu/ciao/).  Spectra are extracted in the [0.6,9.0]
keV band in PI channels, re-binned to have a minimum of 10 counts per
bin and fitted using the XSPEC package (Arnaud 1996).  As background
we use the spectrum extracted from the publicly available background
dataset described in Section~\ref{par:second_unit}.  The
position-dependent RMFs and ARFs are computed and weighted by the
X-ray brightness over the corresponding image region using the
``calcrmf'' and ``calcarf'' tools\footnote{A. Vikhlinin 2000
(http://asc.harvard.edu/ ``Software Exchange'', ``Contributed
Software'').}.  Spectra are fitted with a single temperature absorbed
\textsc{mekal} model using the C-statistics (Cash 1979; Arnaud 1996)
and the Levenberg-Marquardt minimization method (i.e. the XSPEC default 
minimization  method).
In the fit procedure we fix the cluster redshift, metallicity, and
hydrogen column density to the values used as inputs to compute the
\chandra observation.  As result of the fit we obtain what, from now
on, we call projected spectroscopic temperature and its 68 per cent
confidence level error for one interesting parameter, $T_{\rm spec}$
and $\sigma_{\rm spec}$, respectively.

Our goal is to use our \chandra simulator to compare the spectroscopic
temperature $T_{\rm spec}$ to one of the possible temperature
estimator adopted in the analysis of the results of hydro-N-body
simulations.  In particular we use the emission-weighted temperature
$T_{\rm sim}$, which is the one most commonly adopted.  This is
defined as:
\be
T_{\rm sim}\equiv \frac{\int W T dV} {\int W dV}\ .
\ee
In the previous equation, $T$ is the cluster gas temperature, $dV$ is
the volume along the line of sight, while the characteristics of the
emissivity are taken into account by the weight $W$, which is usually
defined as $W=\Lambda(T) \rho^2$, where $\Lambda(T)$ is the cooling
function and $\rho$ the gas density (see, e.g., Navarro, Frenk \&
White 1995).  The emission-weighted temperature for a specific cluster
region is derived directly from the original hydro-N-body simulation
by:
\be
T_{\rm sim}=\frac{\sum_i m_i \rho_i \Lambda(T_i) T_i}{\sum_i m_i \rho_i
\Lambda(T_i)}\ ,
\ee
where the sums are extended to all the particles which are projected
inside the considered region.  To measure the deviation from a
completely isothermal distribution for the gas within the projected
region, we use the relative emission-weighted temperature dispersion,
defined as
\be
 \frac{\sigma_{_T}}{T_{\rm sim}} = \frac{1}{{T_{\rm sim}}}
\left[ 
{\frac{\int W T^2 dV}{\int W dV} - T_{\rm sim}^2}\right]^{1/2} \ :
\label{eq:5}
\ee
quasi-isothermal and highly perturbed regions will have low and high
values of $\sigma_{_T}/T_{\rm sim}$, respectively.

In the following subsections, we will compare $T_{\rm spec}$ to
$T_{\rm sim}$ in two specific cases: i) an ideal toy-model
corresponding to a perfect isothermal cluster, and ii) a more
realistic cluster with a complex thermal structure.

\subsection{Toy isothermal cluster}\label{par:isothermal}

Here we discuss how the spectroscopically inferred temperature $T_{\rm
spec}$ is affected by the total number of detected photons.  At this
goal we take our simulated cluster and set the temperature of all gas
particles to an arbitrarily chosen constant value, $T_{\rm sim}$,
ranging from 3 to 12 keV.  After producing a 300 ks \chandra
observation of this isothermal cluster, we extract all the spectra
from each of the $32^2$ pixels tile regions, defined in
Section~\ref{par:second_unit}.  As the total number of photons in each
spectrum is proportional to the total cluster emissivity in the tile
region where it was extracted from, this simple procedure allows us to
produce a distribution of spectra with the same temperature, but
different total photon counts.  Among the extracted spectra we select
only the ones for which the source flux is higher than the
background. For the exposure time and the fixed region size here
considered, this is equivalent to selecting spectra with net total
counts $N_\gamma> 250$.

Consistently with what already discussed in literature (see, e.g.,
Nousek \& Shue 1989 and references therein), we find a small bias
between the input temperature and the best fit value of the
temperature obtained from spectra with small total number counts
($N_\gamma<10^3$).  This bias increases if the total number counts
gets smaller and/or the temperature is higher.  Nevertheless, we find
that the input temperature value is always consistent within the
1$\sigma$ errors associated with the spectroscopic temperature.

\subsection{Realistic cluster with a complex thermal structure}

To study the effect of temperature inhomogeneities on the final
projected temperature estimates we repeat the analysis of the previous
subsection using the actual photon temperatures obtained from the
simulation, instead of forcing them to be isothermal.  As in
\S~\ref{par:isothermal} we selected only spectra for which the source
flux is higher than the background (net total counts $N_\gamma> 250$).
We estimate $T_{\rm spec}$ and its error $\sigma_{\rm spec}$ by
fitting the data with an absorbed single temperature thermal model.

\begin{figure}
\centering  
 \includegraphics[width=8.8cm]{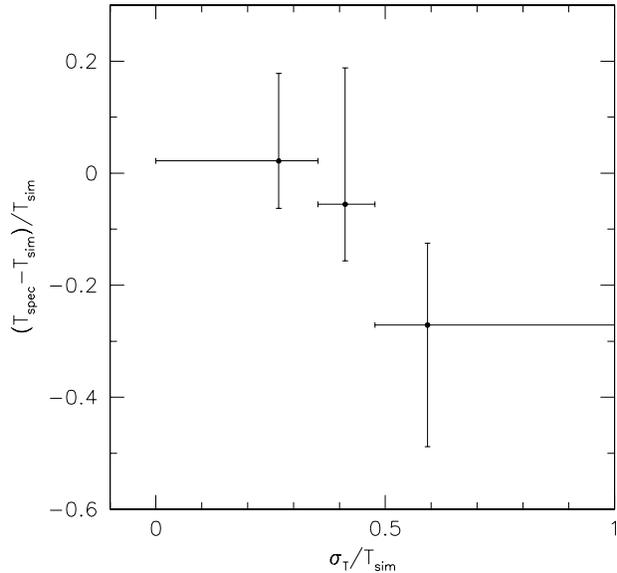}
\caption{
Difference between the spectroscopic and the emission-weighted
temperature estimates ($T_{\rm spec}$ and $T_{\rm sim}$,
respectively), as a function of the degree of thermal inhomogeneity of
the gas, measured by $\sigma_{_T}/T_{\rm sim}$.  The points represent
the median value in each bin and are located in the median value of
$\sigma_{_T}/T_{\rm sim}$.  Vertical error bars indicate the 25 and 75
percentiles of the corresponding distribution inside each bin;
horizontal error bars correspond to the bin size.}
\label{fig:perturb}
\end{figure}

The result of our analysis is shown in Fig.~\ref{fig:perturb}.  All
the considered spectra are divided in three bins with increasing
relative emission-weighted temperature dispersion $\sigma_{_T}/T_{\rm
sim}$ (see equation~\ref{eq:5}).  The three bins, containing the same
number of spectra, correspond to regions that, from the point of view
of the temperature distribution, are highly homogeneous (i.e. almost
isothermal), mildly inhomogeneous and highly inhomogeneous,
respectively.  For each spectrum we calculate the temperature
discrepancy between the spectroscopic and the emission-weighted
estimates, $(T_{\rm spec}-T_{\rm sim})/T_{\rm sim}$.  The points in
the figure represent the median discrepancy in each bin and are
located at the median value of $\sigma_{_T}/T_{\rm sim}$.  The
vertical error bars represent the 25 and 75 percentiles of the
distribution of the discrepancy $(T_{\rm spec}-T_{\rm sim})/T_{\rm
sim}$ inside each bin.  It is interesting to notice that for regions
where the temperature structure is not highly inhomogeneous
(i.e. $\sigma_{_T}/T_{\rm sim}<0.5$), on average we find that $T_{\rm
spec}$ and $T_{\rm sim}$ are consistent, although it is still possible
to obtain discrepancies of the order of 10-15 per cent.  Conversely,
for highly inhomogeneous regions (i.e. $\sigma_{_T}/T_{\rm sim}>0.5$),
we find a larger discrepancy between the spectroscopic and the
emission-weighted temperatures, with on average $T_{\rm spec}$
systematically lower than $T_{\rm sim}$.

Concluding this section, we stress that our results clearly show that,
when strong temperature inhomogeneities are present, the spectroscopic
and the emission-weighted temperature measurements are likely to be
inconsistent with each other.  Consequently, much attention must be
paid in the theoretical interpretation of observational temperatures.
Similar conclusions have been reached by an equivalent analysis of
Mathiesen \& Evrard (2001) in which they compare the overall
spectroscopic and emission-weighted temperatures of a ensemble of 24
simulated clusters of galaxies.  Although with a large scatter, they
claim that the spectroscopic temperature is 20 per cent lower than the
emission-weighted one.

\section{Examples of applications}\label{par:apply}

In this section we discuss, as examples, two different applications of
our simulation method: the computation of temperature profiles and the
production of projected temperature maps of galaxy clusters.

\subsection{Temperature profiles}\label{par:apply1}

\begin{figure*}
{\centering \leavevmode  
\psfig{file=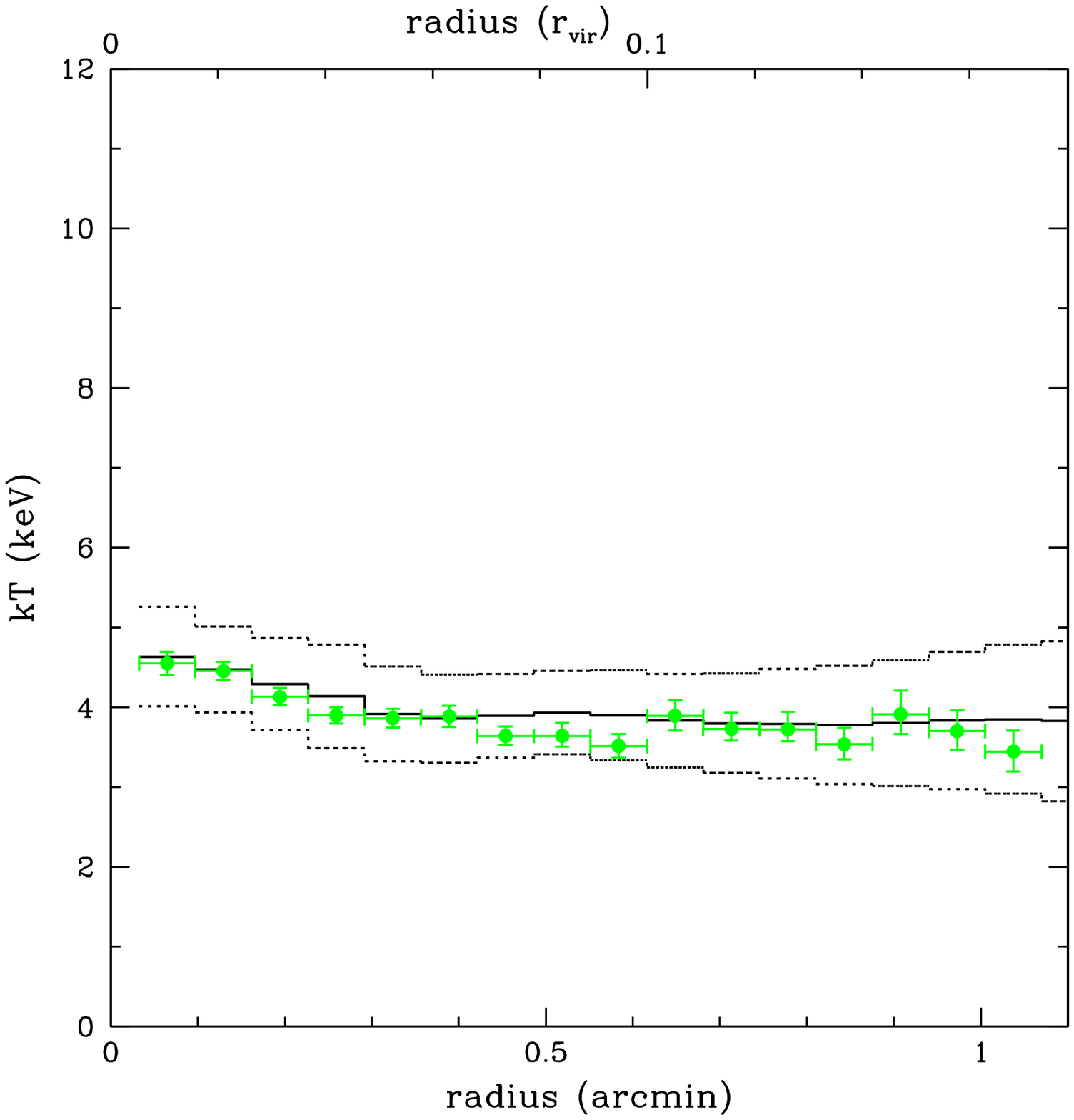,width=.49\textwidth} \hfil  
\psfig{file=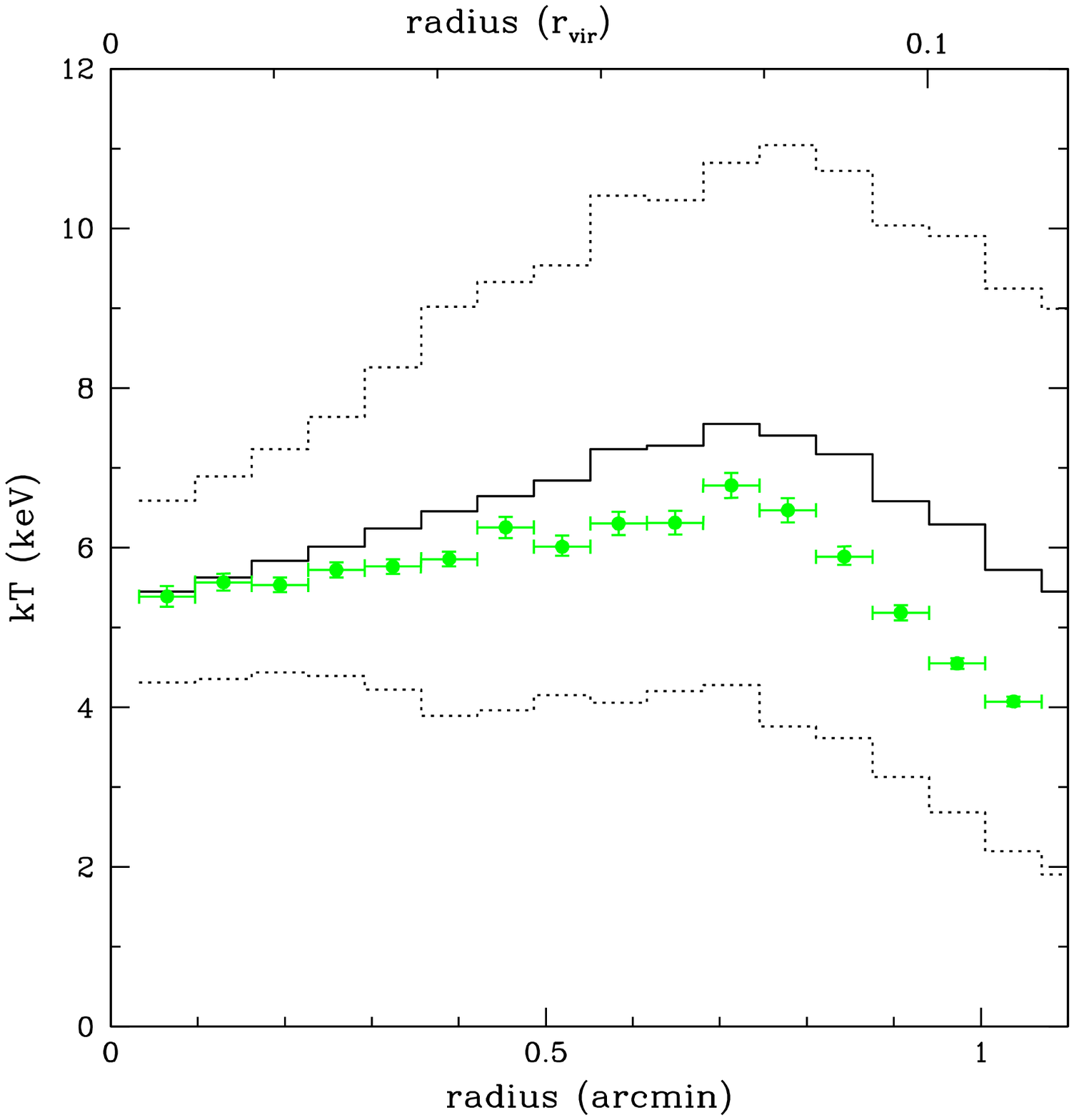,width=.49\textwidth} 
} 
\caption{
Temperature profiles of the galaxy cluster in two different dynamical
phases: a relaxed phase at $z\approx 0.33$ (left panel), and a merging
phase at $z\approx 0.21$ (right panel).  Dots indicate the temperature
profile obtained by fitting the spectra extracted from annular regions
centred on the X-ray peak, after the application of our \chandra
simulator; error bars are at $68$ per cent confidence level for one
interesting parameter.  Solid lines refer to the mean
emission-weighted temperature directly derived from the simulation,
while dotted lines indicate $\sigma_{_T}$, computed as in
equation~\ref{eq:5}.  }
\label{fig:profiles}
\end{figure*}

One of the standard ingredients of the theoretical modeling of galaxy
clusters is the assumption of an isothermal distribution of gas. For
example, this is often used to obtain estimates of their mass by using
the equation of hydrostatic equilibrium. Another standard application
is to directly relate the observed cluster temperature function to the
theoretically estimated mass function to obtain constraints on the
main cosmological parameters, as the matter density parameter and the
normalization and shape of the power spectrum of primordial
fluctuations.

However, the results of high-resolution hydro-N-body simulations
indicate that the temperature shows radial gradients which cannot be
neglected in a dynamical analysis (see, e.g., Evrard, Metzler \&
Navarro 1996; Eke, Navarro \& Frenk 1998; Rasia, Tormen \& Moscardini
2003; Borgani et al. 2003). This is confirmed by recent observational
data, mainly obtained using the X-ray observatories ASCA, BeppoSAX,
and, more recently, \chandra and XMM-Newton (see, e.g., Markevitch et
al. 1998; De Grandi \& Molendi 2002; Pratt \& Arnaud 2002).

In this subsection we use our simulator X-MAS to derive the projected
temperature profiles of both a major merging system and a quasi
relaxed one.  These profiles are compared to the emission-weighted
temperature profiles obtained directly from the hydro-N-body
simulations.  For the merger system we consider the output of the
hydro-N-body simulation at redshift $z=0.21$, already described in
Section~\ref{par:chandra} (see Fig.~\ref{fig:events}).  For the
relaxed system we use the output of the same simulation at $z\approx
0.33$, when the galaxy cluster is not undergoing any strong merging
events.  At this redshift the cluster has a virial mass of $3.87
\times 10^{14} h^{-1} M_\odot$ and a virial radius of $1.7h^{-1}$ Mpc.

To measure the projected temperature profile we extract the spectra
from circular annuli centred on the cluster X-ray peak.  The size of
the bin has been chosen in order to have approximately the same number
of photons inside each annulus.  Again, spectra are extracted in the
[0.3,9.0] keV energy band and fitted with a single temperature
absorbed \textsc{mekal} model with the values for $N_{_H}$,
metallicity, and redshift fixed at the simulated values.  The
spectroscopic temperature profiles $T_{\rm spec}$, together with their
relative 68 per cent confidence level errors $\sigma_{\rm spec}$, are
shown as filled circles in Fig.~\ref{fig:profiles}.  Left and right
panels refer to the relaxed and merging systems, respectively.  In the
same figure we show the emission-weighted temperature $T_{\rm sim}$
and the corresponding value of $\sigma_{_T}$, computed as in
equation~\ref{eq:5}.  As discussed in the previous section, the value
of $\sigma_{_T}$ measures the degree of thermal inhomogeneity of the
considered cluster region. From Fig.~\ref{fig:profiles} we see that,
as expected, the projected radial thermal structure of the cluster in
its relaxed phase is far more homogeneous than the structure in the
perturbed one.  This different degree of thermal homogeneity has
strong implications on the temperature profiles.  In fact, we notice
that while for the relaxed phase the spectral and the emission
weighted temperature profiles are in good agreement, this is not
longer true for the perturbed phase.  Furthermore, we confirm the
systematic trend previously discussed: the spectral temperatures are
lower than the emission-weighted temperatures.

\subsection{Temperature maps}\label{par:apply2}
As a further example of possible applications of our X-ray observatory
simulation package, we now compare the emission-weighted temperature
map determined from the cluster simulation to the projected
temperature map obtained from the spectral analysis of its
corresponding \chandra observation.  In this section we focus only on
the output of the hydro-N-body simulation at $z\approx 0.21$, i.e. the
perturbed phase considered in Section~\ref{par:chandra}.

\begin{figure}
{\centering \leavevmode
\psfig{file=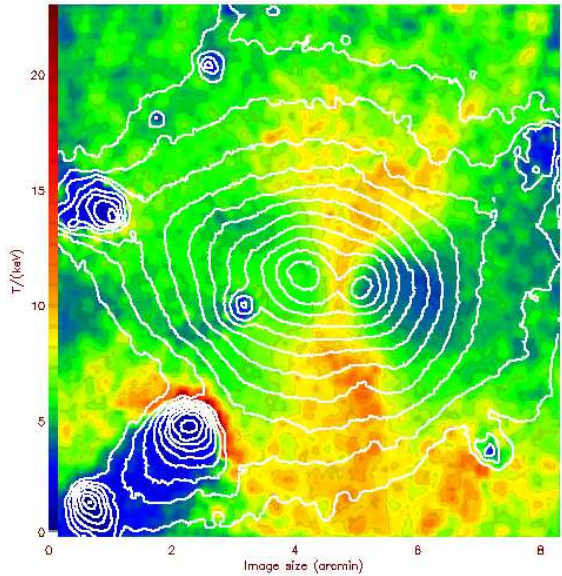,width=.45\textwidth} \hfil
}
\caption{Map of the distribution of the emission-weighted temperature 
for the galaxy cluster at $z\approx 0.21$ as obtained using the gas
particles of the hydro-N-body simulation binned to $1\arcsec$ pixels.
The temperature scale (in keV) is shown on the left.  The contour
levels correspond to the flux distribution shown in
Fig.~\ref{fig:map}.}
\label{fig:com}
\end{figure}

First, we calculate the emission-weighted temperature map of the
simulated clusters.  As the simulation provides us with the density
and the temperature of each particle, the emission-weighted
temperature map for the simulated cluster can be obtained with the
same spatial resolution of its X-ray image. The result is shown in
Fig.~\ref{fig:com}.  For reference, in the same figure we superimpose
the isocontours corresponding to the cluster flux distribution shown
in Fig.~\ref{fig:map}.  As already evident from the temperature
profile (see Fig.~\ref{fig:profiles}), the internal region is far from
being isothermal.  Between the two colder central subclumps with
$T\approx 5-6$ keV, we notice a region with higher gas temperature,
$T\approx 10$ keV, which corresponds to the gas compressed by the
merging blobs. Moreover we notice that the more external merging
subclumps are also significantly colder than the cluster ambient gas.
Of particular interest are the two subclumps on the lower-left corner
of the image: they seem to form a single structure of cold gas. In
addition, we notice the presence of a shock front with a post-shock
gas temperature of approximately 20 keV which is produced by the
motion toward the cluster centre of the most internal of these two
clumps.

To calculate the spectroscopic projected temperature map we use the
300 ks \chandra observation of the simulated cluster described in
Section~\ref{par:chandra}.  We extract spectra using adjacent square
regions.  Over most of the map the extraction regions coincide with
the tile regions defined in Section~\ref{par:second_unit}.  However,
in the outskirt of the cluster, where the surface brightness is lower,
we use larger extraction regions which are obtained by combining two
or more tile regions in such a way that the total net number of
photons per spectrum is $N_\gamma>250$.  Each spectrum is then fitted
with an absorbed single temperature \textsc{mekal} model with
$N_{_H}$, redshift, metallicity fixed to the input values.

\begin{figure*}
{\centering \leavevmode
\psfig{file=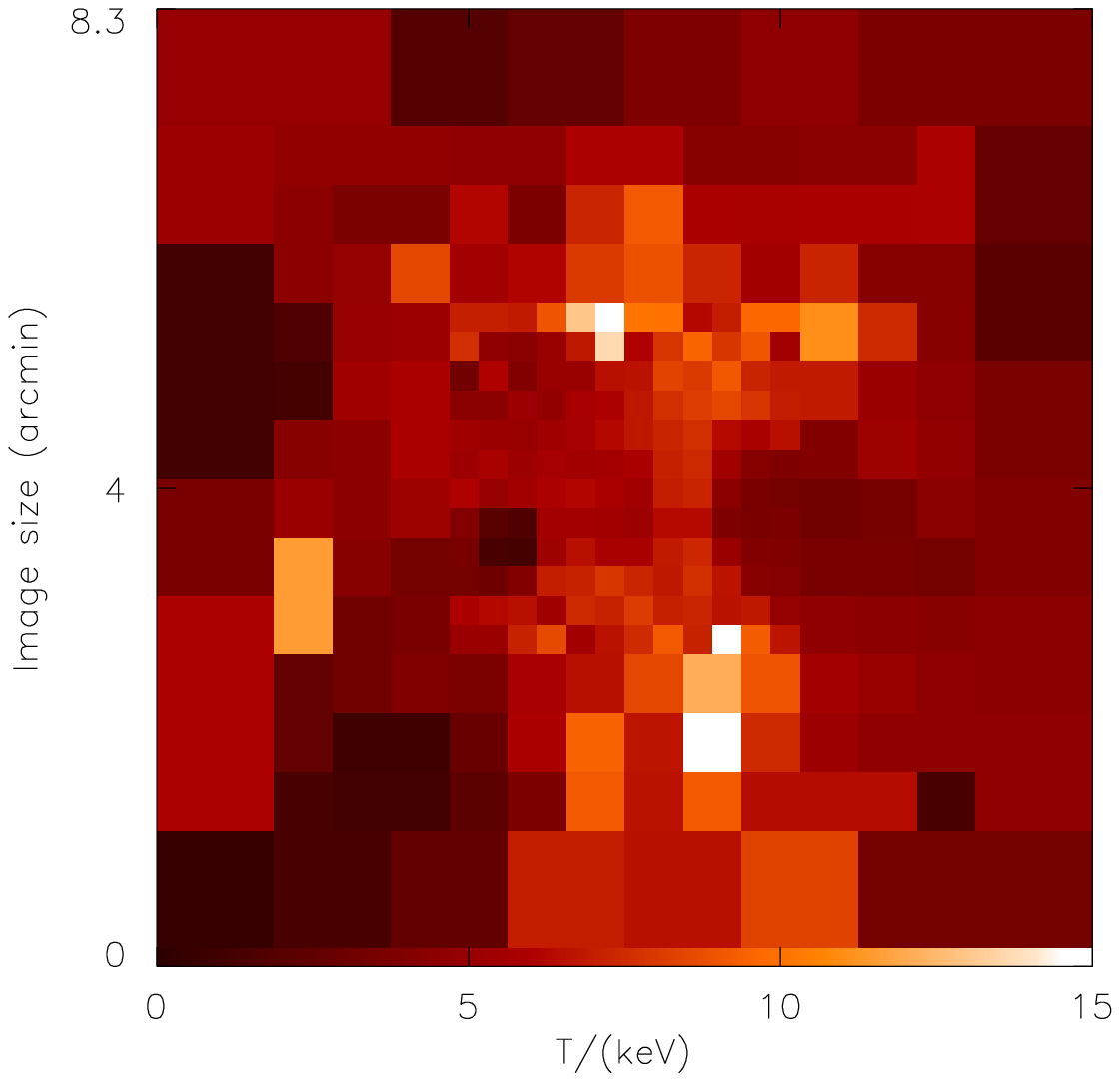,width=.45\textwidth}
\psfig{file=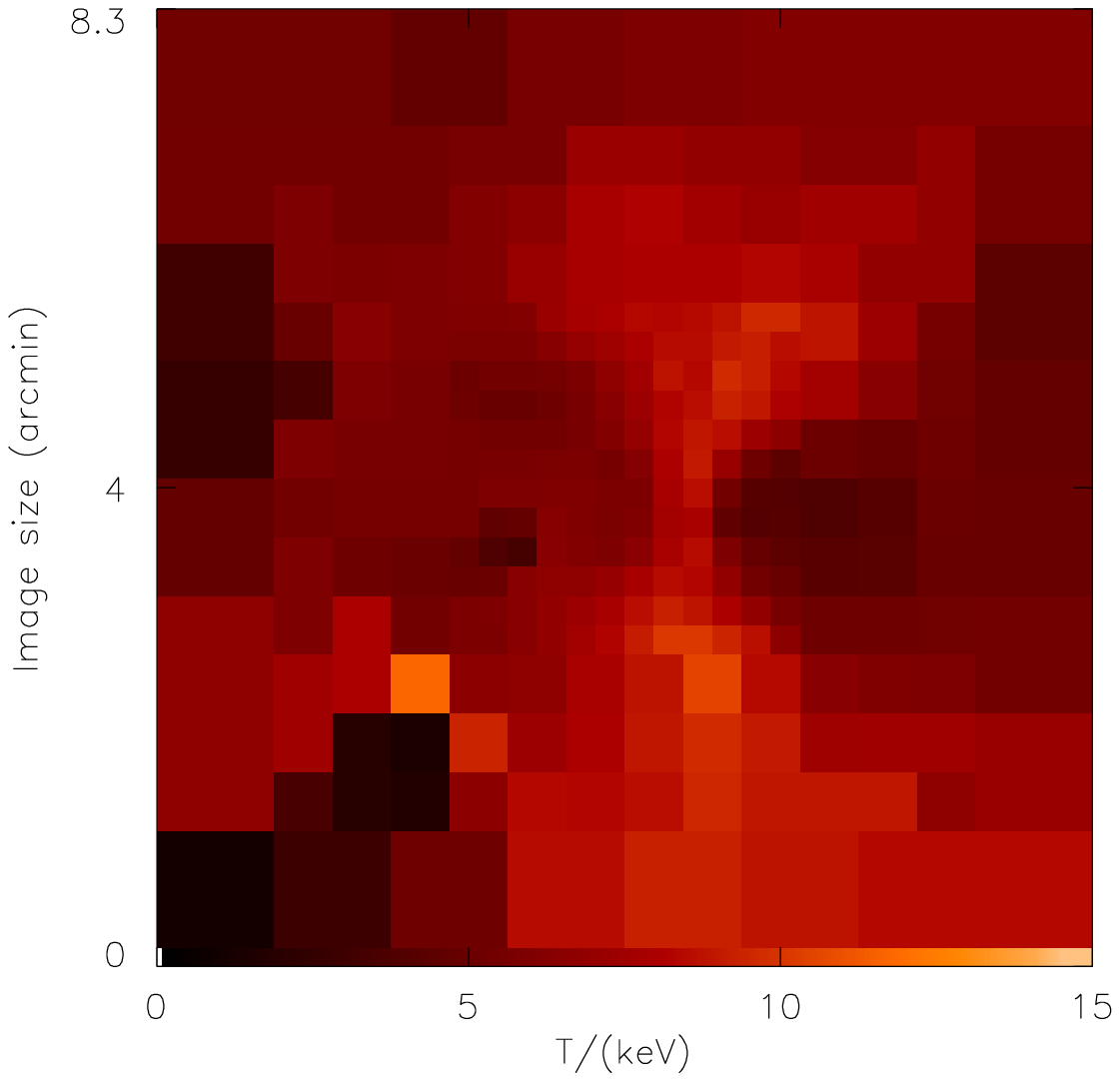,width=.45\textwidth} \hfil
}
\caption{
Temperature maps for the galaxy cluster at $z\approx 0.21$. Left
panel: map of the projected temperature $T_{\rm spec}$ as derived from
the spectroscopic analysis of the \chandra observation.  In the
outskirt of the cluster, larger regions have been considered in order
to have at least 250 net photons per spectrum. Right panel: map of the
emission-weighted temperature $T_{\rm sim}$, shown in
Fig.~\ref{fig:com}, but re-binned as in the left panel.}
\label{fig:comparison2}
\end{figure*}

The projected spectroscopic temperature map $T_{\rm spec}$ is shown in
the left panel of Fig.~\ref{fig:comparison2}. In order to make a
direct comparison of the spectroscopic temperature map to the
emission-weighted one, we degenerate the resolution of the latter to
match the former.  Thus, in the right panel of
Fig.~\ref{fig:comparison2} we report the same emission-weighted
temperature map shown in Fig.~\ref{fig:com}, but re-binned as the
spectroscopic temperature map of the left panel.  It is worth to
notice that, although with a lower spatial resolution, the re-binned
emission-weighted temperature map presents all the main temperature
structures described before. In particular we see that the two central
blobs are cold and we can easily identify the compression-heated gas
between them. At the same way we can recognize the two cold subclumps
on the left-bottom corner, as well as the presence of the shock-heated
gas in front of the innermost of the two which is moving toward the
cluster centre.

If we now compare the emission-weighted temperature map on the right
to the spectral temperature map on the left, we notice that, although
qualitatively similar, they show a number of important differences.
Among others, we point our attention on the fact that: i) the central
merging blobs appear to be colder in the observed spectroscopic
temperature map than in the emission-weighted one; ii) the shock front
produced by the motion of the innermost subclump in the lower-left
corner, clearly visible in the emission-weighted map, is no longer
detected in the spectroscopic temperature map.

\begin{figure*}
{\centering \leavevmode
\psfig{file=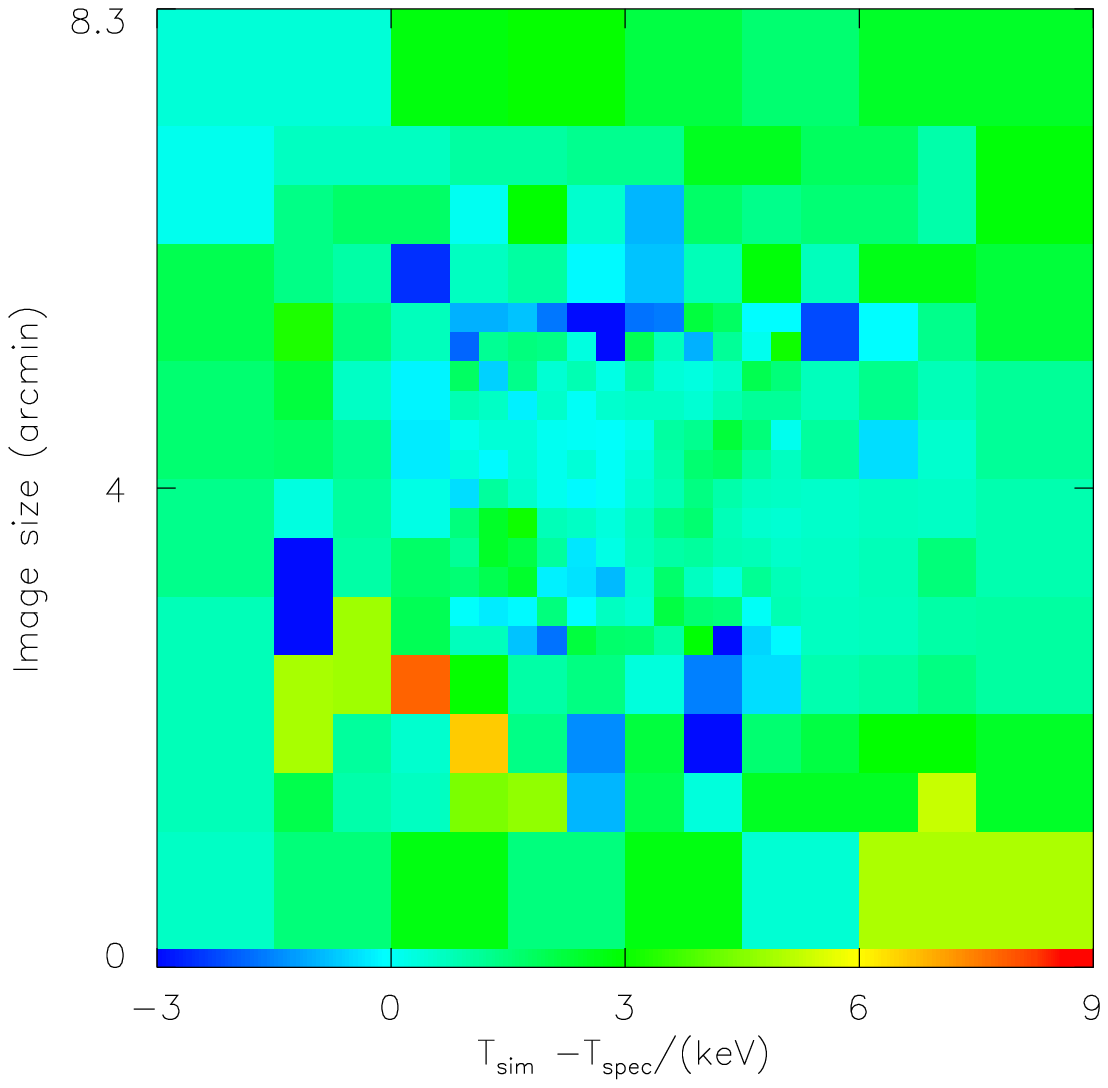,width=.45\textwidth} \hfil
\psfig{file=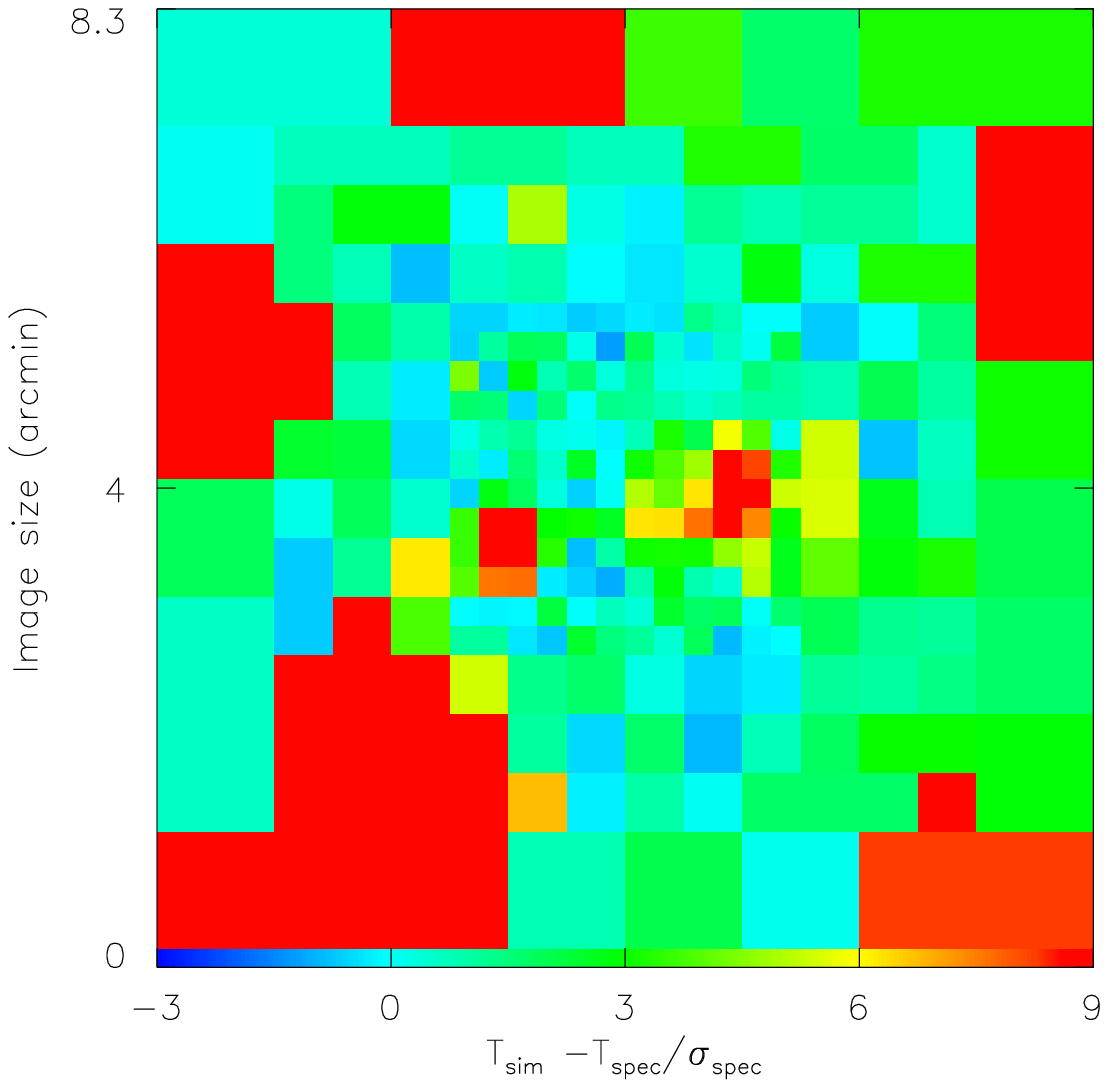,width=.45\textwidth}
}
\caption{
Left panel: map of the differences between the emission-weighted
temperature $T_{\rm sim}$ and the spectroscopic projected temperature
$T_{\rm spec}$ (right and left panels of Fig.~\ref{fig:comparison2},
respectively).  Right panel: significance level of the temperature
differences displayed in the left panel. The map refers to the
quantity $(T_{\rm sim}-T_{\rm spec}) /\sigma_{\rm spec}$, where
$\sigma_{\rm spec}$ is the 68 per cent confidence level error
associated with the spectroscopic temperature measurement.}
\label{fig:comparison3}
\end{figure*}

To better visualize the temperature differences among the two previous
maps, in the left panel of Fig.~\ref{fig:comparison3} we show the
spatial distribution of the difference $T_{\rm sim}-T_{\rm spec}$.  We
find temperature differences $|T_{\rm sim}-T_{\rm spec}|> 1$ keV for
$\ga 50$ per cent of the pixels. In particular we notice that most of
these differences are such that $T_{\rm sim}>T_{\rm spec}$.  To
quantify the significance of this discrepancy, in the right panel of
Fig.~\ref{fig:comparison3} we present the map of $(T_{\rm sim}-T_{\rm
spec})/\sigma_{\rm spec}$ (being $\sigma_{\rm spec}$ the 68 per cent
confidence level error associated with the spectroscopic temperature
measurement).  From this plot we see that most of the temperature
differences are significant at $\ga 3\sigma$ confidence level. In
particular the temperature discrepancies previously noted for the
central subclumps and for the shock regions are significant at $\ga
9\sigma$.

\section{Discussion and Conclusions}\label{par:con}

In this paper we presented a numerical technique and the relative
software package X-MAS devoted to simulate X-ray observations of
galaxy clusters obtained from hydro-N-body simulations. As specific
application, we used this technique to simulate \chandra ACIS-S3
observations.

We stress that one of the main features of our code is that it
generates event files following the same standards used for real
observations.  This is extremely important as it implies that our
simulated observations can be analysed in the same way and with the
same tools of real observations.

Using the results of high-resolution hydro-N-body simulations, we
generated the \chandra observations of a number of simulated
clusters. By performing a standard spectral analysis, we derived the
projected spectral temperature in specific cluster regions.  The
spectral temperature has been finally compared to the
emission-weighted temperature commonly used to describe to cluster gas
thermal properties in numerical works.

Our main finding is that the two temperature estimates are likely to
show a significantly large discrepancy, the spectroscopic temperature
being lower than the emission-weighted temperature. This effect is
more evident if the thermal structure of the cluster within a
particular projected region is relatively complex (i.e. the region is
thermally highly inhomogeneous).  We point out that the data analysis
procedure used in this paper assumes that both the background level
and spectral shape are well known. In fact the background used to
produce the \chandra observations is the same used in the spectral
analysis.  Uncertainties in the background level and spectral shape,
which are quite likely in real observations, would inevitably result
in an increase of the estimated errors and in discrepancies between
the spectroscopic and emission-weighted temperatures even larger than
what has been shown in this paper.  Regardless of the background, the
main reason behind the observed temperature discrepancy can be easily
explained if one considers that spectroscopically the temperature is
determined by fitting a thermal model (in particular its
bremsstrahlung component) to the observed spectrum.  The point is that
the sum of two bremsstrahlung spectra with similar emission but
different temperatures $T_1$ and $T_2$ is no longer a bremsstrahlung
with a given temperature $T_3$.  In fact,
\be
\exp(-E/T_1)+\exp(-E/T_2)\neq A\exp(-E/T_3)\ ,
\ee 
unless $T_1=T_2$.  When such a combined spectrum is fitted with a
single bremsstrahlung spectral model, we obtain a temperature which is
not exactly the mean value of $T_1$ and $T_2$, though it will be an
intermediate value between the two.  The larger is the difference
$|T_2-T_1|$, the bigger will be the discrepancy between the
spectroscopic and the mean temperature.  This result is important
because it implies that much attention must be paid in the theoretical
interpretation of observational temperatures.  Just to give a more
detailed idea of the problem, in Section~\ref{par:apply} we discussed
the implications for two specific cases in which such comparisons have
been done in the past: temperature profiles and temperature maps.

Hydro-N-body simulations have been (and still are) largely used to
derive universal temperature profiles for galaxy clusters. These
profiles are directly compared to the ones obtained from real
observations.  In Section~\ref{par:apply1} we used the \chandra
simulator to compare spectroscopic and emission-weighted temperature
profiles of two extreme phases of the cluster evolution: an almost
relaxed system and a highly perturbed object.  In agreement with what
said before, we find that if the cluster is relaxed, the
emission-weighted temperature profile agrees with the spectroscopic
one.  However, if the system is highly perturbed this is not longer
true.  This result indicate that direct comparisons of observed and
``simulated'' temperature profiles are not fully justified.  In
Section~\ref{par:apply2} we show that a similar problem applies also
to the temperature maps.

In a forthcoming paper we intend to use X-MAS to investigate in detail
a number of problems related to the actual observations of X-ray
galaxy clusters. In particular we intend to study the complex problem
of spectral deprojection and relative cluster mass determination.
Moreover we are planning to use X-MAS to perform observations of
simulated clusters that account for the processes of the metal
enrichment of the ICM.  This will allow us to address the even more
complex problem of determining both the mean and the single-element
metallicity structure of real galaxy clusters.

To conclude we stress that the possible applications of our
simulations of observations are quite vast. Besides the here discussed
problem of the comparisons of outputs of numerical simulations to real
observed galaxy clusters, it can be very useful to better plan
observations with existing X-ray telescopes and even more to verify
what will be the real capabilities of futures ones.  At this goal,
results obtained by using X-MAS will be soon publicly available on the
web.

Concerning our software package, we remind that at present we only
completed the software module that simulates ACIS-S3 \chandra
observations.  Work is in progress to include further modules to
simulate \chandra observations in the ACIS-I mode, and XMM-Newton
observations with both EPIC and MOS detectors.

\section*{Acknowledgments}
This work has been partially supported by Italian MIUR (Grant 2001,
prot. 2001028932, ``Clusters and groups of galaxies: the interplay of
dark and baryonic matter'') and ASI.  We thank also K.A. Arnaud,
S. Borgani, J. McDowell, M. Meneghetti and M. Wise, for clarifying
discussions.  PM, GT and LM are grateful to the Aspen Center for
Physics, where the paper has been discussed and partially written up.

\end{document}